\documentclass[aps,prd,onecolumn,groupedaddress,nofootinbib,preprintnumbers,superscriptaddress]{revtex4}  
\usepackage{graphicx,mathtools,tabu}
\usepackage{epstopdf}
\usepackage{amsmath}
\usepackage{amsfonts}
\usepackage[colorlinks=true,citecolor=red,linkcolor=blue,breaklinks=true]{hyperref}
\usepackage{accents}
\usepackage[figure, table]{hypcap}
\usepackage{amssymb}
\usepackage{appendix}
\usepackage{comment}
\usepackage{bbold}
\usepackage{color}
\usepackage{slashed}
\usepackage{subfigure}
\usepackage{setspace}
\usepackage{footnote}
\usepackage{multirow}
\usepackage{braket}
\usepackage[normalem]{ulem}
\usepackage[utf8]{inputenc}
\usepackage{mathrsfs}
\usepackage{longtable}
\usepackage{enumitem}
\usepackage{booktabs} 
\usepackage{verbatim}

\LTcapwidth=\textwidth

\usepackage{url}
\usepackage{wrapfig}
\usepackage{braket}

 \def\be   {\begin{equation}}  
 \def\ee   {\end{equation}}
 \def\ba   {\begin{array}}     
  \def\ea   {\end{array}}
 \def\bea  {\begin{eqnarray}}  
  \def\eea  {\end{eqnarray}}
 \def\bean {\begin{eqnarray*}}  
 \def\eean {\end{eqnarray*}}

  \def\be {\beta}

\def\to {\rightarrow}

\begin{document}

\preprint{}
\title{Solar neutrinos and $\nu_2$ visible decays to $\nu_1$} 

\author{Andr\'{e} de Gouv\^{e}a}
\email{degouvea@northwestern.edu}
\affiliation{Northwestern University, Department of Physics \& Astronomy, 2145 Sheridan Road, Evanston, IL 60208, USA}
\author{Jean Weill}
\email{jeanweill@u.northwestern.edu}
\affiliation{Northwestern University, Department of Physics \& Astronomy, 2145 Sheridan Road, Evanston, IL 60208, USA}
\author{Manibrata Sen}
\email{manibrata@mpi-hd.mpg.de}
\affiliation{Max-Planck-Institut f\"ur Kernphysik, Saupfercheckweg 1, 69117 Heidelberg, Germany}
\begin{abstract}
Experimental bounds on the neutrino lifetime depend on the nature of the neutrinos and the details of the potentially new physics responsible for neutrino decay.  In the case where the decays involve active neutrinos in the final state, the neutrino masses also qualitatively impact how these manifest themselves experimentally. In order to further understand the impact of nonzero neutrino masses, we explore how observations of solar neutrinos constrain a very simple toy model. We assume that neutrinos are Dirac fermions and there is a new massless scalar that couples to neutrinos such that a heavy neutrino -- $\nu_2$ with mass $m_2$ -- can decay into a lighter neutrino -- $\nu_1$ with mass $m_1$ -- and a massless scalar. We find that the constraints on the new physics coupling depend, sometimes significantly, on the ratio of the daughter-to-parent neutrino masses, and that, for large-enough values of the new physics coupling, the ``dark side'' of the solar neutrino parameter space -- $\sin^2\theta_{12}\sim 0.7$ -- provides a reasonable fit to solar neutrino data. Our results generalize to other neutrino-decay scenarios, including those that mediate $\nu_2\to\nu_1\bar{\nu}_3\nu_3$ when the neutrino mass ordering is inverted mass and $m_2>m_1\gg m_3$, the mass of $\nu_3$.
\end{abstract}

\maketitle

\section{Introduction}

Since the discovery of nonzero, distinct neutrino masses and nontrivial lepton mixing, one can unambiguously conclude that the two heavier neutrinos have finite lifetimes. The weak interactions dictate that these will decay into three lighter neutrinos, assuming the decay is kinematically allowed, or into a lighter neutrino and a photon~\cite{Pal:1981rm}, always kinematically allowed. Quantitatively, however, the weak interactions translate into lifetimes that are many orders of magnitude longer than the age of the universe, exceeding $10^{37}$~years for all values of the neutrino masses and mixing parameters that satisfy existing experimental and observational constraints \cite{ParticleDataGroup:2022pth}. Not surprisingly, the presence of new neutrino interactions and new light states can easily translate into much shorter neutrino lifetimes. 

On the other hand, experimental constraints on the lifetimes of neutrinos -- see, for example, \cite{1984MNRAS.211..277D, Doroshkevich:1989bf, Berezhiani1992, Fogli:1999qt,Choubey:2000an,Lindner:2001fx,Beacom:2002cb,Joshipura:2002fb,Bandyopadhyay:2002qg, Beacom:2002vi,Beacom:2004yd,Berryman:2014qha,Picoreti:2015ika,FRIEMAN1988115,Mirizzi:2007jd,GonzalezGarcia:2008ru,Maltoni:2008jr,Baerwald:2012kc,Broggini:2012df,Dorame:2013lka,Gomes:2014yua,Abrahao:2015rba,Coloma:2017zpg,Gago:2017zzy,Choubey:2018cfz,Chianese:2018luo,deSalas:2018kri, deGouvea:2019goq,Funcke:2019grs,Escudero:2020ped,Abdullahi:2020rge,Akita:2021hqn,Picoreti:2021yct,DeGouvea:2020ang, Chen:2022idm,deGouvea:2022cmo} -- are absurdly far from the expectations of the standard model plus massive neutrinos. These rely on experiments with neutrinos that travel long distances before they are detected, ranging from earth bound reactor and accelerator neutrino experiments (1 to 1,000~km), solar neutrino experiments (500~light-seconds), neutrinos from SN1987A (170,000 light-years), to indirect inferences regarding the properties of the cosmic neutrino background. All experimental bounds on the neutrino lifetime are model dependent. They depend on the nature of the neutrinos -- are neutrinos Majorana fermions or Dirac fermions? -- the decay mode -- are there visible particles, such as neutrinos or photons, in the final state? --  and the dynamics of the interaction responsible for the decay -- does it involve left-chiral or right-chiral  neutrino fields? Furthermore, as we explored in \cite{deGouvea:2022cmo}, in the case where the decay involves active neutrinos in the final state, the neutrino masses qualitatively impact the neutrino decay and how it manifests itself experimentally.

Solar neutrinos provide robust, reliable bounds on the neutrino lifetime. Given everything we know about neutrino masses and neutrino mixing, the solar neutrino spectrum is well known and, it turns out, it is characterized by an incoherent mixture of the neutrino mass eigenstates, so the impact of neutrino decay is easy to visualize. There is also a wealth of solar neutrino data collected in the last several decades. Here we will concentrate on data from SuperKamiokande~\cite{Super-Kamiokande:1998zvz} and SNO~\cite{SNO:2009uok,SNO:2011hxd} -- on $^8$B neutrinos -- and on data from Borexino ~\cite{BOREXINO:2018ohr} -- on $^7$Be neutrinos --  in order to explore how observations of solar neutrinos constrain a very simple toy model, taking finite neutrino masses into account. We assume that neutrinos are Dirac fermions and there is a new massless scalar that couples to neutrinos such that a heavy neutrino can decay into a lighter neutrino and a massless scalar. We find that the constraints on the new physics coupling depend, sometimes significantly, on the ratio of the daughter-to-parent neutrino masses, and that, for specific values of the new physics coupling, the ``dark side'' of the solar neutrino parameter space \cite{deGouvea:2000pqg} provides a reasonable fit to solar neutrino data. We also find that ``high-energy'' solar neutrino data complement the data on ``low-energy'' solar neutrinos in a very impactful manner. 

In Section~\ref{sec:model}, we discuss the model under investigation and the characteristics of the neutrino decay processes mediated by the model. In Section~\ref{sec:analysis}, we briefly summarize the effects of neutrino decay on neutrino flavor evolution, highlighting solar neutrinos. We discuss the different experimental data and constraints in Sections~\ref{sec:Borexino}, \ref{sec:superk}, and \ref{sec:SNO} while combined results are presented in Section~\ref{sec:combined}. Section~\ref{sec:conclusions} contains a summary of our findings along with generalizations and some parting thoughts. 

\section{The model}
\label{sec:model}
We assume the neutrino mass eigenstates $\nu_i$ with mass $m_i$, $i=1,2$, interact with a massless scalar boson $\varphi$ via the following Lagrangian,
\begin{equation}
    \mathcal{L}\supset g\,\bar{\nu}_1 \mathbb{P_L} \nu_2 \varphi\, + H.c.,
\label{eq:Lagrangian}
\end{equation}
where the neutrinos are Dirac fermions and $\mathbb{P_L}$ is the left-chiral projection operator. Neutrino mass eigenstates are defined in the usual way, $m_2>m_1$, and we do not consider similar interactions involving $\nu_3$. This operator mediates the decay of a $\nu_2$ into a $\nu_1$. The analysis of this decay in the case $m_1=0$  was performed, e.g., in~\cite{Berryman:2014qha}, where it was argued that a subset of solar neutrino data, as well as KamLAND data, can be used to constrain the invisible decays of $\nu_2$: $m_2 \Gamma_2 < 9.3\times10^{-13}\,{\rm eV}^2$. A consistent but more precise bound was later obtained by the SNO collaboration \cite{SNO:2018pvg}. We expect different results once the daughter neutrinos have non-zero masses. Other consequences of Eq.~(\ref{eq:Lagrangian}) will be briefly discussed in Sec.~\ref{sec:conclusions}.

\begin{figure}[!t]
       \centering \includegraphics[width=0.6\textwidth]{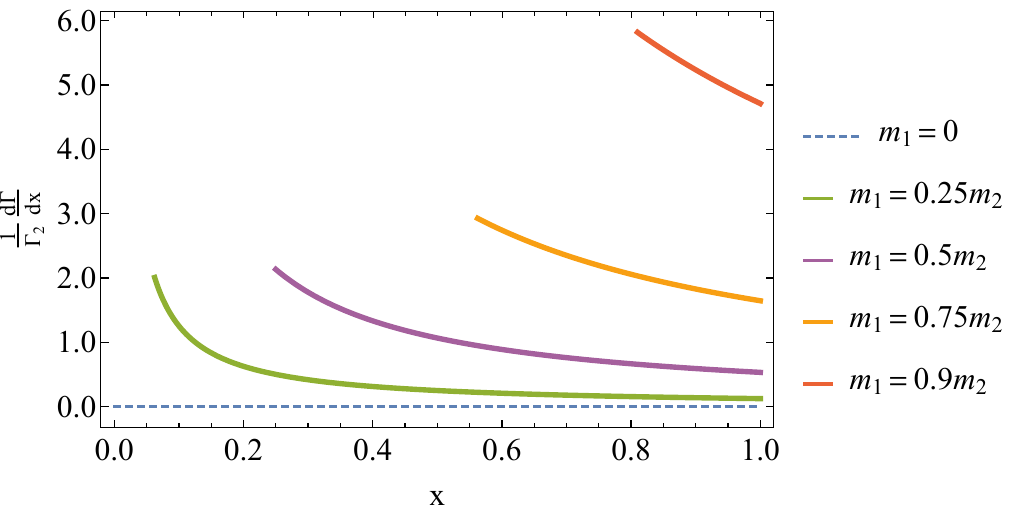}
    \caption{Differential decay distribution, normalized to the total width $\Gamma_2$, as a function of  $x=E_1/E_2$, for a left-handed helicity $\nu_2$ with laboratory-frame energy $E_2$ decaying into a massless scalar and a left-handed helicity $\nu_1$ with laboratory-frame energy $E_1$, assuming neutrino decay is governed by Eq.~(\ref{eq:Lagrangian}). The different curves correspond to different values of $m_1/m_2$.}
\label{fig:diffGamma} 
\end{figure}

Eq.~(\ref{eq:Lagrangian}) only contains the right-chiral component of the $\nu_1$ field. In the limit $m_1\to 0$, in a $\nu_2$ decay process, only right-handed helicity $\nu_1$ are produced,\footnote{For the same reason, only left-handed helicity $\bar{\nu}_1$ are produced when a $\bar{\nu}_2$ decays.} independent from the polarization state of the $\nu_2$. For all practical purposes, right-handed helicity neutrinos are inert and cannot be detected. For $m_1\neq 0$, there is a nonzero probability for the production of left-handed helicity -- hence detectable -- daughter $\nu_1$. This probability grows as the daughter neutrino mass approaches the parent neutrino mass. In this limit,  in fact, the decay into left-handed helicity $\nu_1$ dominates over the decay into right-handed $\nu_1$. Fig.\,\ref{fig:diffGamma} depicts the differential decay width of a left-handed helicity $\nu_2$ with energy $E_2$ into a left-handed helicity $\nu_1$ with energy $E_1$, normalized to the total decay width, as a function of  $x=E_1/E_2$, for different values of $m_1/m_2$. Clearly, as $m_1\rightarrow m_2$, there is a significant increase in the contribution of the helicity preserving -- left-handed daughter -- channel. Furthermore, as $m_1\to m_2$, the decay spectrum is compressed; energy-momentum conservation implies that the heavy daughter inherits most of the parent energy while the massless $\varphi$ comes out with only a tiny fraction of the allowed energy. For many more details and discussions, see \cite{deGouvea:2022cmo}. 

The neutrino-decay physics mediated by Eq.~(\ref{eq:Lagrangian}) is governed by three parameters: the dimensionless coupling $g$ and the neutrino masses $m_1$ and $m_2$. The difference of the neutrino masses squared $\Delta m^2_{21}$ is experimentally well constrained, mostly by the KamLAND reactor neutrino experiment \cite{KamLAND:2008dgz}, so we use $g$ and the ratio of the neutrino masses $m_1/m_2$ to define the remaining two-dimensional parameter space of the model.  The decay width of a $\nu_2$ at rest multiplied by its mass,\footnote{Experiments directly constrain $m_2\Gamma_2$; the laboratory-frame decay width is $m_2\Gamma_2/E_2$.} is~\cite{deGouvea:2022cmo}
\begin{equation}
\label{eq:mGD}
m_2\Gamma_2 = g^2\frac{\Delta m^2_{21}}{32\pi}\left(1+\frac{m_1^2}{m_2^2}\right).
\end{equation} 
For fixed $\Delta m^2_{21}$, $m_2\Gamma_2$ depends linearly on $g^2$ and only very weakly on the ratio of the neutrino masses, varying by a factor of two as the value of  $m_1/m_2$ covers its entire allowed range from zero to one. There is, however, an experimental upper bound to $m_1/m_2$. It is trivial to compute
\begin{equation}
	\label{eq:mfixed}
	m_2^2=\frac{\Delta m^2_{21}}{\left(1-\frac{m_1^2}{m_2^2}\right)},~~~~~
	m_1^2=\frac{\Delta m^2_{21}\left(\frac{m_1^2}{m_2^2}\right)}{\left(1-\frac{m_1^2}{m_2^2}\right)},
\end{equation}
and note that, for a fixed $\Delta m^2_{21}$, both the values of $m_1$ and $m_2$ diverge as $m_1/m_2\rightarrow 1$. Nonetheless $m_1/m_2$ values very close to one are experimentally allowed. Consider, for example, $\Delta m^2_{21}=7.54\times 10^{-5}~{\rm eV}^2$ and an upper bound of $0.1~\rm{eV}$ for $m_2$.  Using Eq.~(\ref{eq:mfixed}), this upper bound translates into $m_1/m_2\le 0.996$. On the other hand, arbitrarily small values of $m_1/m_2$ are allowed as long as the neutrino mass ordering is normal ($m_3>m_2>m_1$). For the inverted neutrino mass ordering ($m_2>m_1>m_3$), $m_1/m_2\ge0.985$. In summary, virtually all values of $m_1/m_2$ are allowed by the data, including values very close to one. In the case of the inverted neutrino mass ordering, only values of $m_1/m_2$ close to one are allowed.

\section{Analysis strategy}
\label{sec:analysis}
In vacuum, allowing for the possibility that $\nu_2$ with helicity $r$ and energy $E_h$ can decay into a $\nu_1$ with helicity $s$ and energy $E_l$ with associated partial differential decay width $d\Gamma_{rs}/dE_l$, the differential probability (per unit $E_l$) for a $\nu_{\alpha}$ with helicity $r$ and energy $E_h$ to behave as a $\nu_{\beta}$ with helicity $s$ and energy $E_l$ after it has traveled a distance $L$ is~\cite{Lindner:2001fx,Coloma:2017zpg} 
\begin{eqnarray}
\label{eqn:vis}
   \frac{dP_{{\nu^r_\alpha\rightarrow\nu^s_\beta}}(L)}{dE_l}&=&
   \left|\sum_{i=1}^3 U_{\alpha i}\,U_{\beta i}^*\, {\rm exp}\left(-i\frac{m_i^2L}{2E_h}\right){\rm exp}\left(-\delta_{i2}\frac{m_2\Gamma_2\,L}{2E_h}\right)\right|^2\, \delta(E_h-E_l) \, \delta_{rs}\nonumber\\
   &+& 
  \frac{1}{\Gamma_2} \frac{d\Gamma_{rs}}{dE_l}\vert U_{\alpha 2}\vert^2\vert U_{\beta 1}\vert^2\left[1-{\rm exp}\left(-\frac{m_2\Gamma_2 L}{E_h}\right)\right]\,,
\end{eqnarray}
where $U_{\alpha i}$, $\alpha = e, \mu, \tau$, $i=1,2,3 $ are the elements of the leptonic mixing matrix and $\Gamma_2$ is the $\nu_2$ total decay width. The first term encodes the contribution from the surviving parent neutrino, including oscillations, while the second term includes the contribution from the daughter neutrino. 

Solar neutrinos, instead, are well described as incoherent mixtures of the mass eigenstates. Hence, the initial state produced inside the sun with energy $E_h$ exits the sun as a $\nu_i$ with probability $P_i(E_h) $, $i=1,2,3$, and all neutrinos are left-handed ($r=-1$). The differential probability that the neutrino arriving at the earth with energy $E_l$ is potentially detected as a $\nu_{\beta}$ with helicity $s$ is 
\begin{eqnarray}
	\label{eqn:sun}
	\frac{dP_{{\nu_{\odot}\rightarrow\nu^s_\beta}}(L)}{dE_l}&=&\left[P_1(E_h) |U_{\beta1}|^2+P_2(E_h) |U_{\beta 2}|^2\exp\left(-\frac{m_2\Gamma_2\,L}{E_h}\right)+P_3(E_h)|U_{\beta 3}|^2\right] \delta_{-1s}\delta(E_h-E_l) \nonumber \\
	&+&   \frac{1}{\Gamma_2} \frac{d\Gamma_{-1s}}{dE_l}P_2(E_h)\vert U_{\beta 1}\vert^2\left[1-{\rm exp}\left(-\frac{m_2\Gamma_2 L}{E_h}\right)\right]\,.
\end{eqnarray}
The impact of the decay is as follows. The $\nu_2$ population decays exponentially and is, instead, replaced by a $\nu_1$ population with a softer energy spectrum and with positive and negative helicities. Furthermore, the daughter energy spectrum is also distorted relative to the parent one by the energy dependency of the exponential decay; higher energy parents decay more slowly than lower energy ones.

It is pertinent to make a few comments regarding the $\nu_3$ component of the solar neutrino flux. $P_3\sim 0.02$ for all $E_h$ of interest so the original $\nu_3$ contribution to the flux is very small. Had we allowed for interactions involving $\nu_3$, these would not lead to especially interesting effects for solar neutrinos. In more detail, if the neutrino mass ordering were normal ($m_3>m_2>m_1$), the new interaction involving $\nu_3$ would mediate potentially visible $\nu_3$ decays.  In this case, however, the impact of the decay-daughter population -- equivalent to the second line in Eq.~(\ref{eqn:sun}) -- would be suppressed by $P_3$ and hence small relative to the dominant $\nu_2$ and $\nu_1$ original populations. Instead, if the neutrino mass ordering were inverted  ($m_2>m_1>m_3$), the new interaction involving $\nu_3$ would mediate potentially visible $\nu_2$  and $\nu_1$ decays into $\nu_3$. In this case, at least when it comes to detectors predominantly sensitive to the $\nu_e$ component of the beam, the daughter population would be almost invisible since $|U_{e3}|^2\sim 0.02$ is very small relative to $|U_{e1}|^2,|U_{e2}|^2$.

The differential number of events at a detector that is sensitive to $\nu_{\beta}$ via the weak interactions, including visible decays, is~\cite{Lindner:2001fx} 
\begin{equation}
\label{eqn:Ntot}
    \frac{d^2N_{\nu_\odot\rightarrow\nu_{\beta}}(L)}{d\tilde{E}_l\,dE_l}=\sum_{s=-1,1}R(\tilde{E}_l,E_l)\sigma_s(E_l)\int^{E_{\rm max}}_{E_l} dE_h \Phi(E_h)\frac{dP^{\rm visible}_{\nu_\odot\rightarrow \nu^s_\beta}(L)}{dE_l},
\end{equation}
where $\Phi(E_h)$ denotes the neutrino energy spectrum at production and $E_{\rm max}=E_l m^2_2/m^2_1$ is the kinematical upper bound on $E_l$. The resolution function connecting the true energy $E_l$ and the detected energy $\tilde{E}_l$ is $R(\tilde{E}_l,E_l)$. The total cross-section for detecting a $\nu_{\beta}$ wih helicity $s$ is $\sigma_s(E_l)$. For right-handed helicity neutrinos, $s=1$, the weak cross-section is suppressed by $m_1^2/E_l^2$, and is set to zero throughout.


\section{Simulations and Results}

Here, we consider in turn the solar neutrino data from Borexino, Super-Kamiokande, and SNO, and estimate their sensitivity to visible solar neutrino decays. When simulating event rates at Borexino, we considered 1,072 days of Borexino Phase-II data taking~\cite{BOREXINO:2018ohr}. For Super-Kamiokande, we consider 504 days of data taking, corresponding to Super-Kamiokande Phase I~\cite{Super-Kamiokande:1998zvz}, and for SNO, we consider 365 days of data taking, which corresponds to roughly the first two phases of SNO \cite{SNO:2011hxd}.

We make use of the PDG parameterization for the elements of the mixing matrix and, when applicable, use the following values for the oscillation parameters of interest \cite{ParticleDataGroup:2022pth}:
\begin{equation}
    \begin{array}{c}
       \sin^2\theta_{12}=0.307; \,\,  \sin^2\theta_{13}=0.0218; \,\, 
       \Delta m_{21}^2=7.54\times 10^{-5}{\rm eV}^2; \,\, \Delta m_{31}^2=2.47\times 10^{-3}{\rm eV}^2\,.
   \end{array}
   \label{eq:osc_param}
\end{equation} 
Throughout, our main goal is to understand the impact of the daughter neutrino mass $m_1$ and explore whether nontrivial neutrino decays allow for a different fit to the solar neutrino data.  

\subsection{Borexino}
\label{sec:Borexino}

Borexino~\cite{BOREXINO:2018ohr} is a 280~ton liquid scintillator detector located underground at the Laboratori Nazionali del Gran Sasso (LGNS) in Italy. Its main focus is the detection of solar neutrinos, in particular $^{7}$Be neutrinos, through neutrino--electron scattering. Neutrinos are detected via the scintillation light which is emitted isotropically during the propagation of the recoil electron and detected by 2212 photo-multiplier tubes, allowing for the measurement of the recoil-electron energy. When simulating event rates at Borexino, we considered 1,072 days of Borexino Phase-II data taking, $N_{\rm{tar}}=3\times 10^{31}$ targets coming from the 100 tons of fiducial mass. We approximated the $^{7}$Be neutrino differential energy flux by a delta function. The kinematical parameter most relevant to the experiment is the electron recoil energy, which follows a continuous distribution governed by the neutrino--electron scattering process. 

\begin{figure}[htb]
\centering
\includegraphics[width=0.6\textwidth]{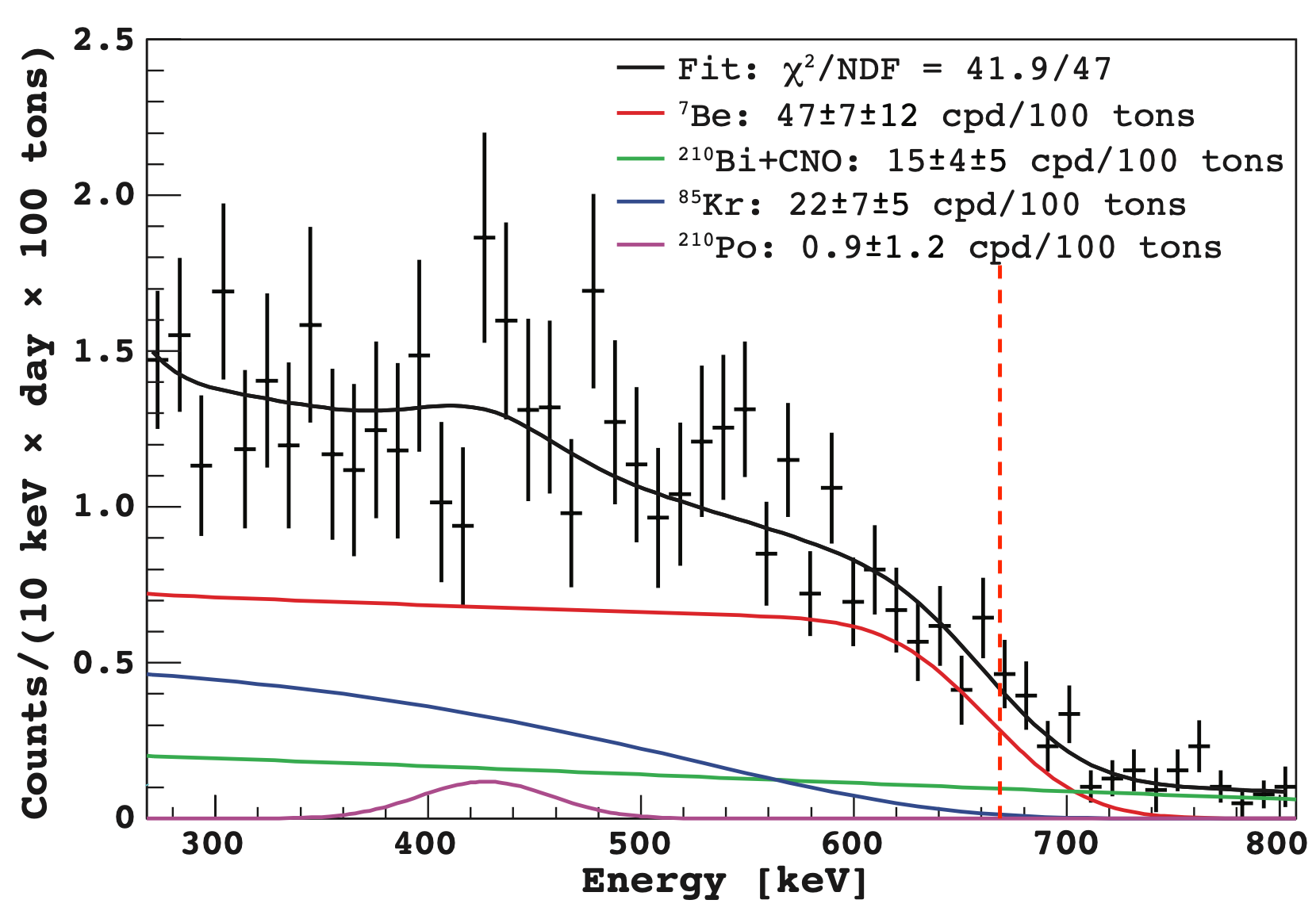}
\caption{\label{fig:borex_collab} From \cite{Borexino:2007kvk}. Observed number of events per 10~keV, per day, per 100~tons reported by the Borexino Collaboration, as a function of the recoil-electron kinetic energy. The different curves correspond to the results of a fit, performed by the Borexino Collaboration, to the different physics processes that contribute to data, as labelled. The vertical dashed line indicates the kinematical upper bound for the scattering of $^7$Be neutrinos with electrons at rest.}
\end{figure}

The experiment succeeds at detecting $^7$Be neutrinos by achieving the strictest radio-purity levels. A detailed understanding of the main backgrounds was therefore necessary to properly estimate the sensitivity of Borexino to neutrino decays. Fig.\,\ref{fig:borex_collab}, from \cite{BOREXINO:2018ohr}, depicts the main backgrounds for the solar neutrino measurement. These come from radioactive processes involving $^{210}$Bi, $^{85}$Kr, and $^{210}$Po ~\cite{BOREXINO:2018ohr}. In our analyses, we treat the different background components independently.  Using Fig.\,\ref{fig:borex_collab}, we fit for the shape of the different background components, which we hold fixed.

For different values of the decay and mixing parameters $g,r\equiv m_1/m_2$, and $\sin^2\theta_{12}$ we compute the equivalent of the red curve in Fig.\,\ref{fig:borex_collab}. We simplified our analyses by considering a Gaussian energy resolution function for the $^{7}$Be spectrum and assuming $100\%$ efficiency. We restricted our analyses to recoil energies between $200~\rm{keV}$ and $665~\rm{keV}$.  665~keV is the maximum kinetic energy of the recoil electron for $862~\rm{keV}$ $^{7}$Be neutrinos. For higher recoil energies, we did not have enough information on the Borexino energy resolution in order to perform a trustworthy analysis and decided, conservatively, to exclude these data points from the analysis. The $665~\rm{keV}$ threshold is highlighted in Fig.\,\ref{fig:borex_collab} with a red vertical dashed line. 

We bin both the background and signal curves in order to perform a $\chi^2$ fit to the data in Fig.\,\ref{fig:borex_collab}. The value of the unoscillated $^{7}$Be flux, which is rather well known, is held fixed. We first analyze the data assuming neutrinos are stable ($g=0$) and fit for the normalization of each background component along with that of the $^7$Be neutrino contribution. We further constrain the $^{210}$Bi background by including the data associated to recoil kinetic energy bins between $740~\rm{keV}$ and $800~\rm{keV}$, making the simplifying assumption that only $^{210}$Bi events contribute inside that window. Having done that, henceforth we fix the normalization of the different background components to these extracted best-fit values. 

\begin{figure}[htb]
\centering
\includegraphics[width=0.4\textwidth]{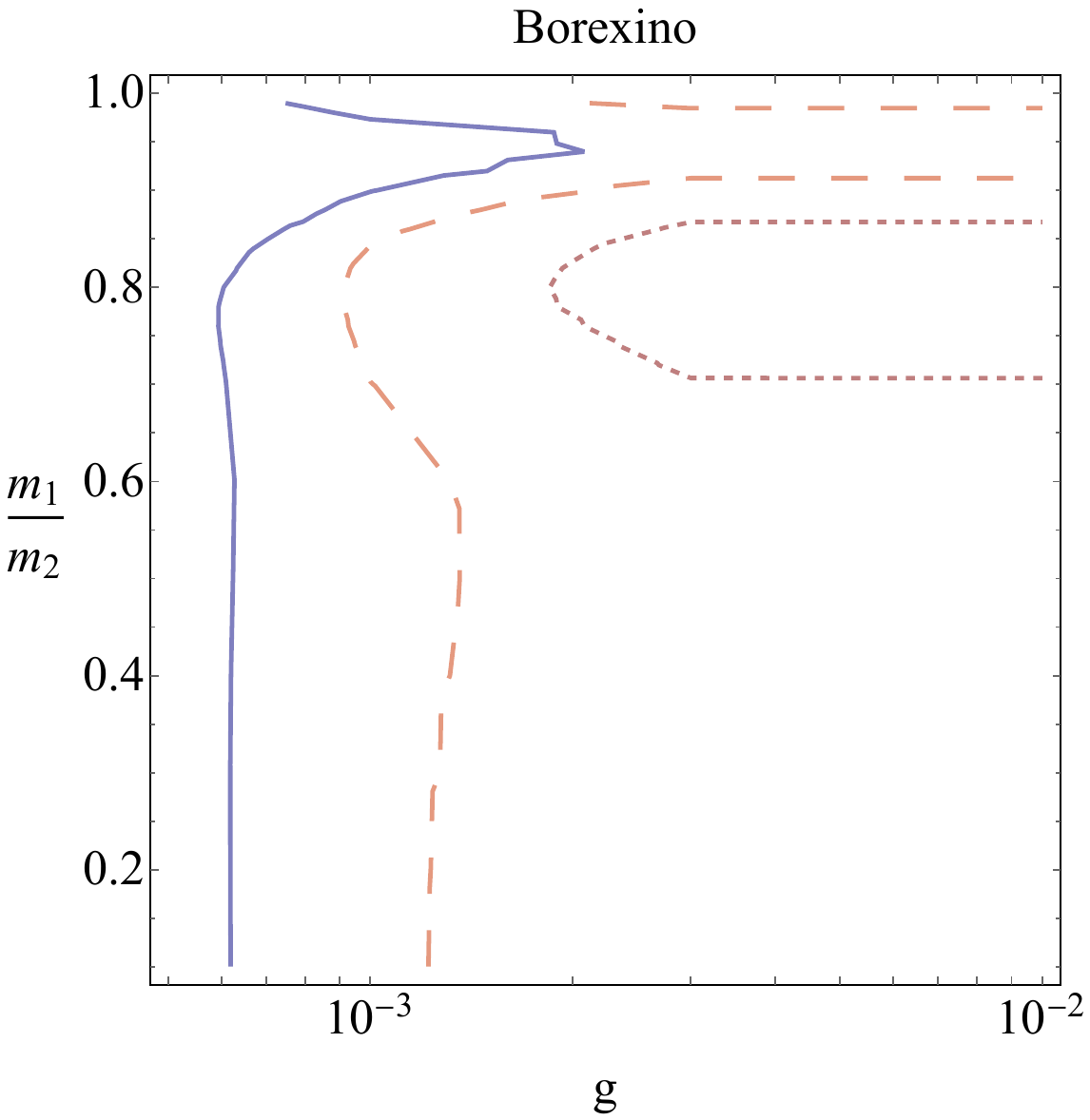}~
\includegraphics[width=0.4\textwidth]{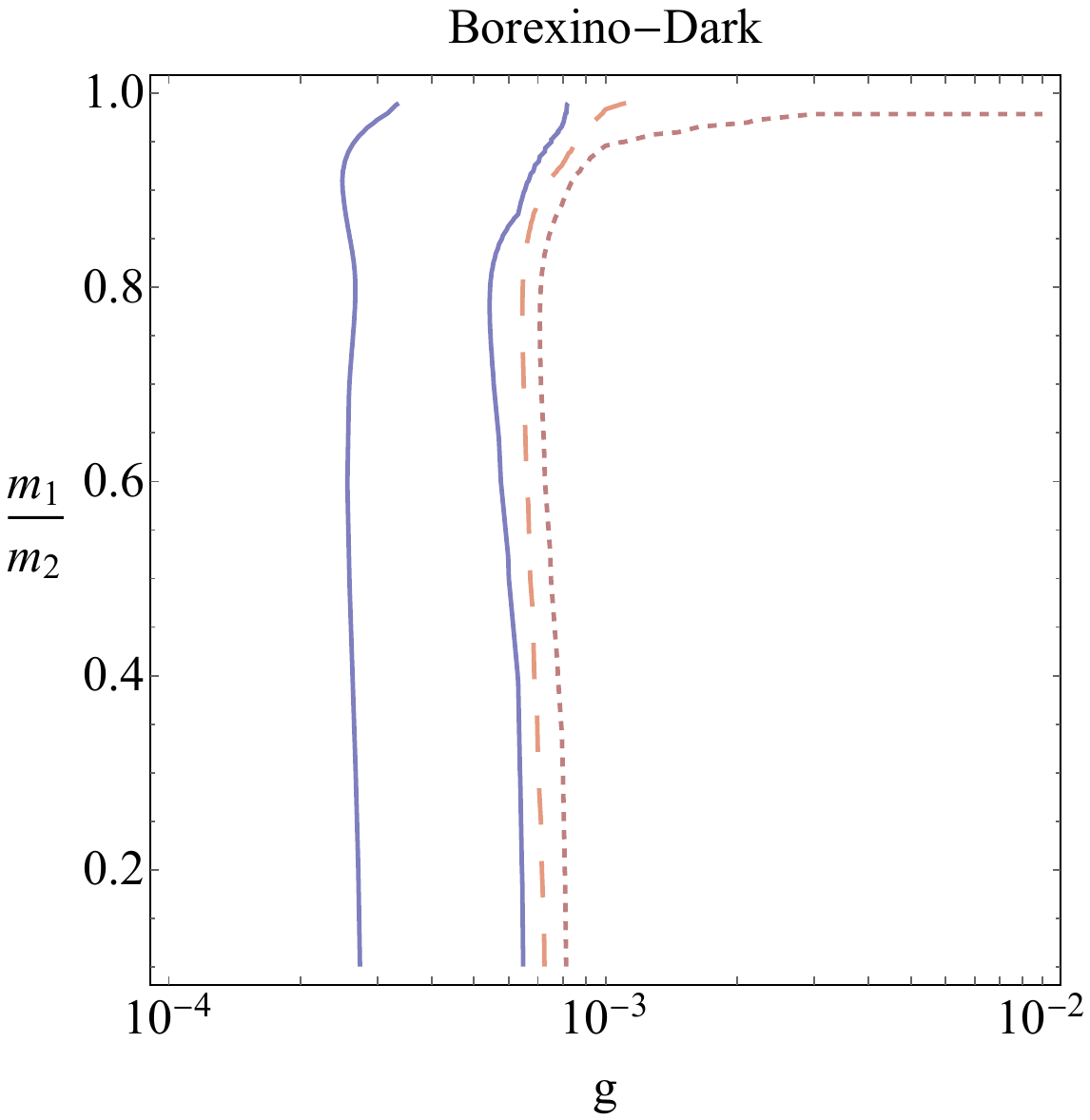}
\caption{\label{fig:Sinsq3} Regions of the $g\times r$, $r=m_1/m_2$, parameter space allowed by Borexino data assuming that external data constrain $\sin^2\theta_{12}=0.30\pm0.05$ (left) or $\sin^2\theta_{12}=0.70\pm0.05$ (right). See text for the details. The different contours correspond to one $\sigma$ or $\Delta \chi^2=2.30$ (solid), two $\sigma$ or $\Delta \chi^2=6.18$ (big dashed), and three $\sigma$ or $\Delta \chi^2=11.83$ (small dashed).}
\end{figure}

Taking all of this into account, we compute $\chi^2(g,r,\sin^2\theta_{12})$, find $\chi^2_{\rm min}$, the minimum value of $\chi^2$, and define the boundaries of `allowed' and `excluded' regions of parameter space using fixed values of $\Delta \chi^2\equiv\chi^2-\chi^2_{\rm min}$. In our analyses, we marginalize over the value of $\sin^2\theta_{12}$ and add a Gaussian prior in order to include external constraints on this mixing angle. We first make use of the following prior: $\sin^2\theta_{12}=0.30\pm 0.05$, selected from the current best fit value for $\sin^2\theta_{12}$ and consistent with the uncertainty reported by KamLAND \cite{KamLAND:2008dgz}. Fig.\,\ref{fig:Sinsq3}(left) depicts the regions of the $g\times r$ parameter space allowed at the one-, two-, and three-sigma levels ($\Delta \chi^2=2.30$, $6.18$, and 11.83, respectively).
 
Using the results from Sec.~\ref{sec:analysis}, and taking into account that matter effects are small for $^7$Be solar neutrino energies, the electron neutrino survival probability for $^7$Be neutrinos, integrating over the daughter neutrino energy, is well approximated by\footnote{For clarity, all approximate expressions here assume $\sin^2\theta_{13}=0$, unless otherwise noted.} 
\begin{equation}
\label{eqn:PSimp}
    P_{\rm{ee}}=\cos^4\theta_{12}+\sin^4\theta_{12}e^{-m_2\Gamma_2 L/E_h}+\left(1-e^{-m_2\Gamma_2 L/E_h}\right)\times f(r,\sin^2\theta_{12}),
\end{equation}
where $f(r,\sin^2\theta_{12})$ is a function of $r$ and $\sin^2\theta_{12}$.  The first two terms correspond to the contribution of the surviving parents while the last term comes from the visible daughter component. The function $f(r,\sin^2\theta_{12})$, while relatively cumbersome, has the following simple limit: $f(r\rightarrow 0,\sin^2\theta_{12})= 0$ for all $\sin^2\theta_{12}$. This limit follows from the fact that, as the daughter mass $m_1\to 0$, all daughters have right-handed helicity and are hence invisible. $f(r,\sin^2\theta_{12})$ also has an approximate upper limit, which we will discuss momentarily.  

When $m_2\Gamma_2 L/E_h$ is small, the decay effects are not significant. As discussed earlier,  for fixed $\Delta m^2_{21}$, $m_2\Gamma_2$ depends exclusively, for all practical purposes, on $g$. For the Earth--Sun distance and $^7$Be neutrino energies, $m_2\Gamma_2 L/E_h\ll 1$ for $g\lesssim 0.001$. In this region, the electron neutrino survival probability is 
 \begin{equation}
\label{eqn:PInv}
    P_{\rm{ee}}=P^{\rm{No Decay}}_{\rm{ee}}=\cos^4\theta_{12}+\sin^4\theta_{12}.
\end{equation}
This limiting case is apparent in the left panel in Fig.\,\ref{fig:Sinsq3} where all values of $g\lesssim 0.001$ are allowed, mostly  independent from $r$.

In the opposite regime -- $m_2\Gamma_2 L/E_h\gg1$ -- Eq.~(\ref{eqn:PSimp}) simplifies to 
\begin{equation}
\label{eqn:PVis}
    P_{\rm{ee}}=\cos^4\theta_{12}+f(r,\sin^2\theta_{12}), 
\end{equation}
again keeping in mind that matter effects are small for $^7$Be neutrino energies. In this region of the parameter space, the electron neutrino survival probability depends on $r$ but does not significantly depend on $g$. This behavior is apparent in Fig.\,\ref{fig:Sinsq3}(left) where the contours become horizontal lines. The behavior of $\Delta \chi^2$ is governed by two effects: the ``missing'' $\nu_2$-component of the parent population and the behavior of the visible daughter contribution. The effect of the missing $\nu_2$-component can be seen when $r\ll 1$ and $f$ is very small. The fact that $P_{ee}<P_{ee}^{\rm{No Decay}}$ allows one to disfavor that region of the parameter space. For larger values of $r$, $f$ is finite and the daughter contribution can make up for the missing $\nu_2$ component of the flux, as is apparent in Fig.\,\ref{fig:Sinsq3}(left). More quantitatively, when the decays are prompt relative to the Earth--Sun distance, the daughter contribution is of order $\sin^2\theta_{12}\cos^2\theta_{12}\times$Br(visible), where Br(visible) is the probability that the daughter from the decay has left-handed helicity and is therefore visible. Numerically, the combination $\sin^2\theta_{12}\cos^2\theta_{12}\sim 0.2$ and, for Br(visible)$\sim 0.5$, it turns out that $\sin^4\theta_{12}\sim \sin^2\theta_{12}\cos^2\theta_{12}\times$Br(visible). Note that for $r\to 1$, Br(visible)$\to 1$  and the decay solution ``overshoots'' the no-decay electron-neutrino survival probability, a behavior that is also reflected in the left panel in Fig.\,\ref{fig:Sinsq3}. Finally, we highlight that values of $r\sim 0.8$ are slightly more disfavored relative to other values of $r$ in the limit where the decay is prompt. The reason is partially related to the distortion of the daughter neutrino energy spectrum relative to the parent one concurrent with a significant fraction of visible decays. 

\subsubsection{Dark Side}

If one ignores solar neutrino data, all detailed information on $\sin^2\theta_{12}$ comes from reactor antineutrino experiments. In fact, until the JUNO experiment \cite{JUNO:2015zny} starts collecting and analyzing data, all detailed information comes from the KamLAND experiment. The experimental conditions are such that, to an excellent approximation, KamLAND is only sensitive to $\sin^22\theta_{12}=4\sin^2\theta_{12}\cos^2\theta_{12}$ and cannot distinguish $\theta_{12}$ from $\pi/2-\theta_{12}$.\footnote{Given the solar constraints we are investigating here, $\nu_2$ decays are irrelevant at KamLAND:  the lifetimes of interest are way too long relative to the ${\cal O}(100~\rm km)$ KamLAND baselines. } $^8$B solar neutrino data break the degeneracy and rule out the so-called dark side of the parameter space, $\sin^2\theta_{12}>0.5$. Since we are introducing a hypothesis that modifies the flavor evolution of solar neutrinos, we investigate the constraints on $g$ and $r$ restricting $\theta_{12}$ to the dark side. 

In the absence of oscillations, because matter effects are negligible for $^7$Be solar neutrino energies, Borexino cannot rule out the dark side of the parameter space: for both $\sin^2\theta_{12}=0.3$ and $\sin^2\theta_{12}=0.7$, $P^{\rm{No Decay}}_{\rm{ee}}=0.58$. This does not hold when the parent neutrino is allowed to decay. We repeat the exercise discussed earlier in this section with the prior $\sin^2\theta_{12}=0.7\pm 0.05$ and depict our results in the right panel in Fig.\,\ref{fig:Sinsq3}. Comparing to the left panel in Fig.\,\ref{fig:Sinsq3}, it is clear that the exchange symmetry is broken. Dark and light side priors on $\sin^2\theta_{12}$ lead to different results for both the invisible and visible contribution, leading to significant differences between the hypotheses.

As before, when $g \lesssim 0.001$ decay effects are irrelevant -- the lifetime is too long -- and the Borexino data are not sensitive to $g$ or $r$. When $g\gtrsim 0.001$, the results obtained with the two different priors differ considerably. These differences are simplest to analyze qualitatively when $r$ is small. In this limiting case, the daughter neutrinos are effectively invisible and the $\nu_2$ component of the flux has enough time to completely disappear. We are left with 
\begin{equation}
P_{ee}\simeq\cos^4\theta_{12}.
\end{equation}
In the light side, as discussed earlier, $\sin^2\theta_{12}=0.3$ translates into $P_{ee}\sim 0.5$, not too far from the central value preferred by the no-decay scenario, $P_{ee}=0.58$. Instead, in the dark side, $\cos^2\theta_{12}=0.3$ and $P_{ee}\sim 0.1$, markedly smaller than the the preferred value in the no decay scenario. This is apparent when comparing the two panels in Fig.\,\ref{fig:Sinsq3}; the dark-side constraints are stronger than the light-side ones and the large $g$, small $r$ region is excluded, in the dark side, at more than the three-sigma level.\footnote{In the dark side, for relatively small values of $g$, one runs into a region of parameter space that is slightly preferred over the no-decay hypothesis. This preference is not statistically significant.}  

For larger values of $r$, the daughter contribution improves the quality of the fit in the dark side when the $\nu_2$ decay is prompt. Following the discussion below Eq.~(\ref{eqn:PVis}), the daughter contribution cannot exceed approximately 0.2 and, in the dark side assuming the decay hypothesis, when $g\gtrsim 0.001$, $P_{ee}< 0.3$, always less than the  central value preferred by the no-decay scenario ($P_{ee}^{\rm No Decay}=0.58$). Nonetheless, the region of parameter space for $r\to 1$ and large values of $g$ is disfavored at less than the three-sigma level.

\subsection{Super-Kamiokande}
\label{sec:superk}
Super-Kamiokande (SK) is a 50~kton water Cherenkov detector running in Japan. Solar neutrinos interact inside the water mainly through neutrino--electron scattering, which is sensitive to neutrinos of all flavors. The scattered electron produces Cherenkov radiation in the water, which can be detected. To simulate events in SK, we compute the neutrino--electron cross-section and consider 504 days of data taking ~\cite{Super-Kamiokande:1998zvz}, consistent with phase I of SK solar neutrino data. SK has collected over 5,000 days of solar neutrino data but we restrict our analyses to phase I for a couple of reasons. First, the detector changed qualitatively after phase I and our simulations below are best matched to the results presented in \cite{Super-Kamiokande:1998zvz}. Second, we are most interested in exploring the differences between Borexino, SK, and SNO (discussed in the next subsection) and how these different data sets complement one another, as opposed to how strict are the solar constraints on our neutrino decay hypothesis. For this reason, we benefit most from combining data sets that are roughly of the same size. The number of targets is taken to be $N_{\rm tar}=3\times10^{33}$. The resolution function is taken to be $\mathcal{S}(E_{\rm tr})=-0.084 + 0.376 \sqrt{E_{\rm tr}} + 0.040 E_{\rm tr}$. 

SK is sensitive to $^8$B neutrinos. The neutrino oscillation parameters are such that $^8$B neutrinos are predominantly $\nu_2$: $P_2\gtrsim 0.9$ for the energy range of interest \cite{Nunokawa:2006ms} and, in the absence of neutrino decays, the electron neutrino survival probability is $P_{ee}\sim |U_{e2}|^2$. SK data are consistent with $|U_{e2}|^2\equiv \sin^2\theta_{12}\cos^2\theta_{13}\sim 0.3$. Since $\cos^2\theta_{13}\sim 0.98$, this translates into $\sin^2\theta_{12}\sim0.3$. As discussed in detail earlier, invisible $\nu_2$ decays lead to a suppression of the measured flux. This suppression can, in principle, be compensated by increasing the value of $\sin^2\theta_{12}$. 

Similar to the Borexino discussion, here it is also easy to estimate the impact of the invisible decays. If a fraction $\epsilon$ of the $\nu_2$ population survives, $P_{ee}\sim (1-P_2)\cos^2\theta_{12}+P_2\epsilon\sin^2\theta_{12}$, ignoring $\sin^2\theta_{13}$ effects. If epsilon is not zero, one could tolerate a population drop of up to almost 70\% by jacking up $\sin^2\theta_{12}$ all the way to one. On the other hand, when $\epsilon=0$, because $P_2\gtrsim 0.9$, it is impossible to lower $\sin^2\theta_{12}$ enough to reach $P_{ee}\sim 0.3$ even in the limit $\cos^2\theta_{12}\to1$. 
Of course, KamLAND data constrain $\sin^2\theta_{12}\sim 0.3$. Hence, the combination of reactor and solar data prevent one from varying $\sin^2\theta_{12}$ with complete impunity. KamLAND data are, however, also completely consistent with $\sin^2\theta_{12}\sim0.7$ -- the dark side solution discussed in the last subsection -- allowing some extra flexibility. 

The situation is different when the daughters of the decay are visible, something we expect when $r\equiv m_1/m_2$ is large. In this case, the daughters contribute to the number of events in SK, contributing an amount of order $P_2\cos^2\theta_{12}(1-\epsilon)$. Including the parent contribution, $P_{ee}\sim \cos^2\theta_{12}-\epsilon P_2\cos2\theta_{12}$. Now, if all $\nu_2$ decay ($\epsilon\to 0$), $P_{ee}\sim \cos^2\theta_{12}$ which may provide a good fit to SK data if $\theta_{12}$ is in the dark side. 

\begin{figure}[!t]
 \centering
    \begin{tabular}{ccc}
        \includegraphics[width=0.32\textwidth]{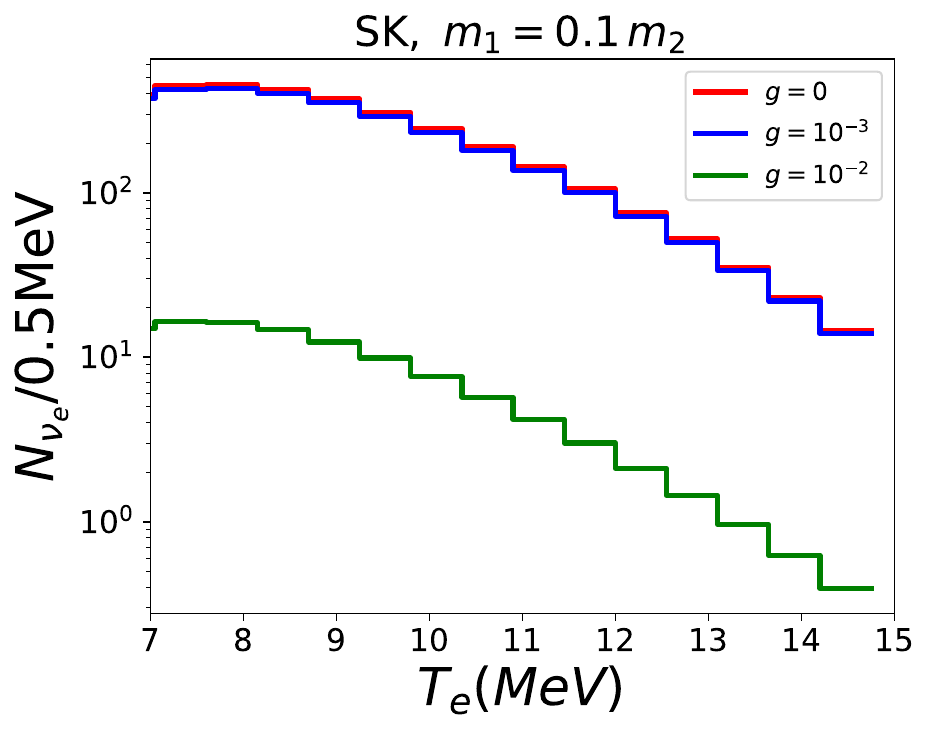}~
        \includegraphics[width=0.32\textwidth]{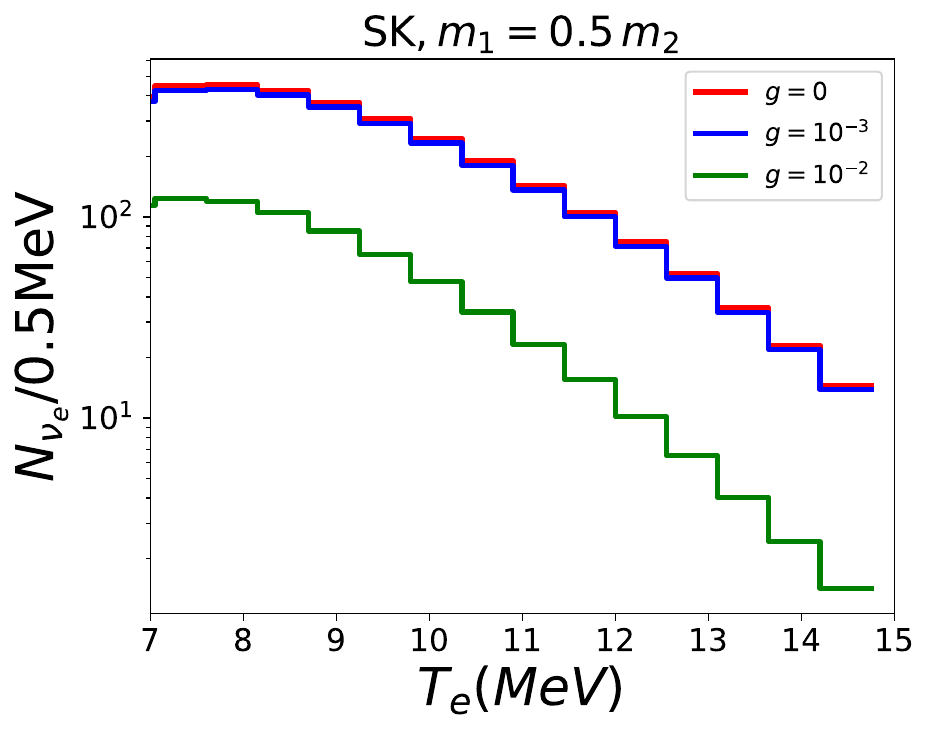}~
        \includegraphics[width=0.32\textwidth]{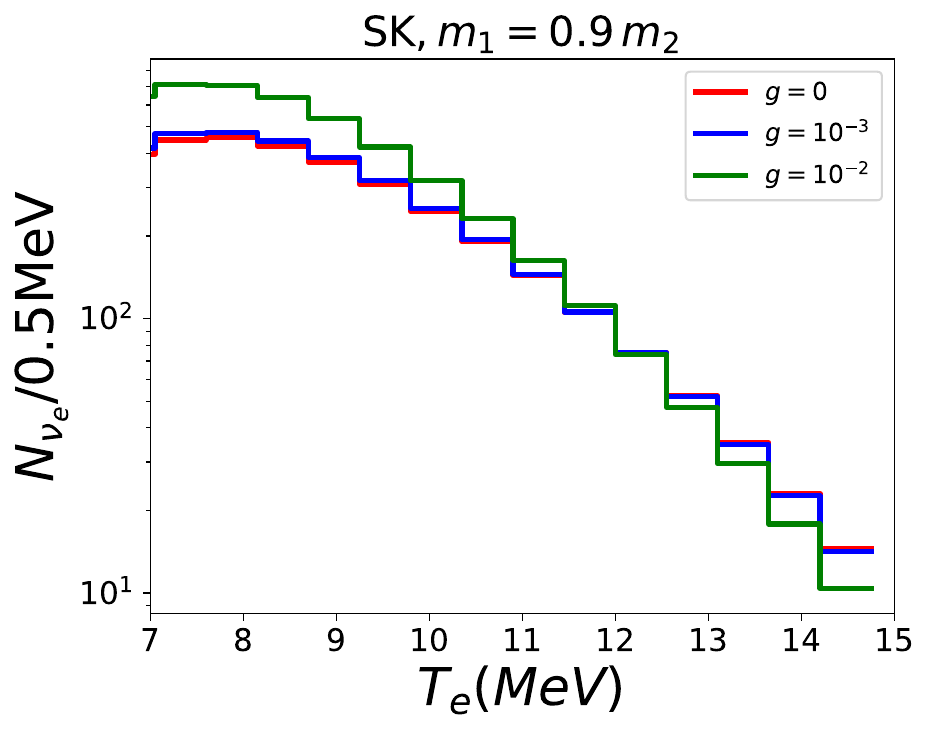}
    \end{tabular}
    \caption{
    \label{fig:SK_visdec_md} Simulated Super-Kamiokande data for different values of the neutrino decay coupling $g$ and the ratio of the parent to daughter neutrino mass, $r=m_1/m_2$: $r=0.1$ (left), $r=0.5$ (center), and $r=0.9$ (right). $g=0$ corresponds to stable neutrinos, the standard oscillation scenario. All oscillation parameters are set to the values listed in Eq.~(\ref{eq:osc_param}). }
\end{figure}

The discussion in the preceding paragraphs is meant to be qualitative and we now turn to a more quantitative estimate of SK's sensitivity. Fig.\,\ref{fig:SK_visdec_md} shows the expected number of events as a function of the recoil-electron kinetic energy for different values of $g$ and $r$. Here, $\Delta m^2_{21}=7.54\times 10^{-5}{\rm eV}^2$, $\sin^2\theta_{12}=0.307$, and $\sin^2\theta_{13}=0.0218$. We find that for $r=0.1$ (left-hand panel), as the value of the coupling $g$ increases, the number of events decreases. In this regime, most of the decays are invisible and the impact of the nonzero daughter mass is negligible. However, for larger values of $r$, the visible contribution is significant. For $r=0.9$ (right-panel) the impact of the visible daughters is strong enough that one observes an increase of the expected number of events as the coupling $g$ increases. One should also note that the recoil-electron kinetic energy spectrum is distorted relative to the no-decay hypothesis. 

\begin{figure}[!h]
 \centering
        \includegraphics[width=0.4\textwidth]{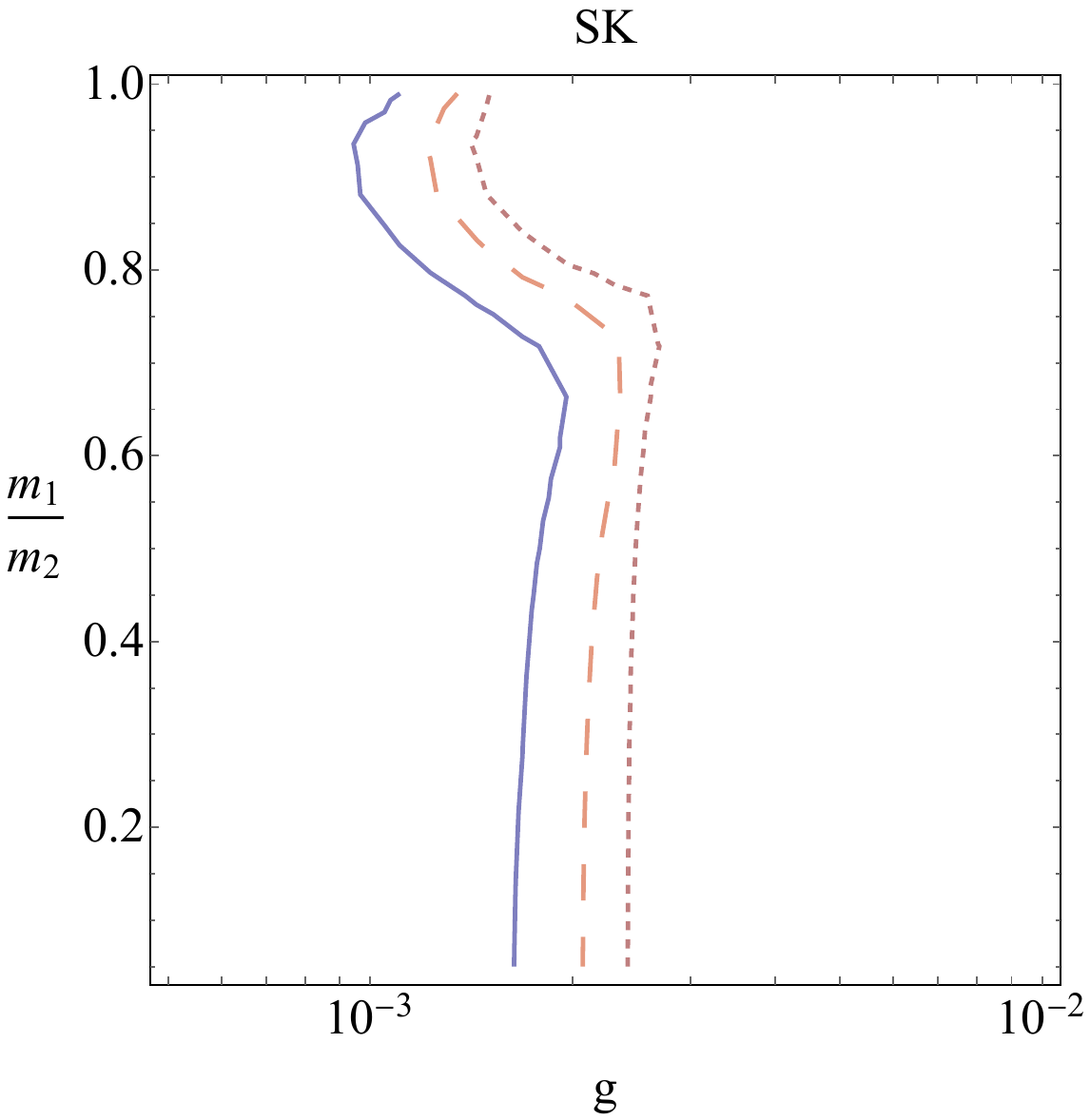}~
         \includegraphics[width=0.4\textwidth]{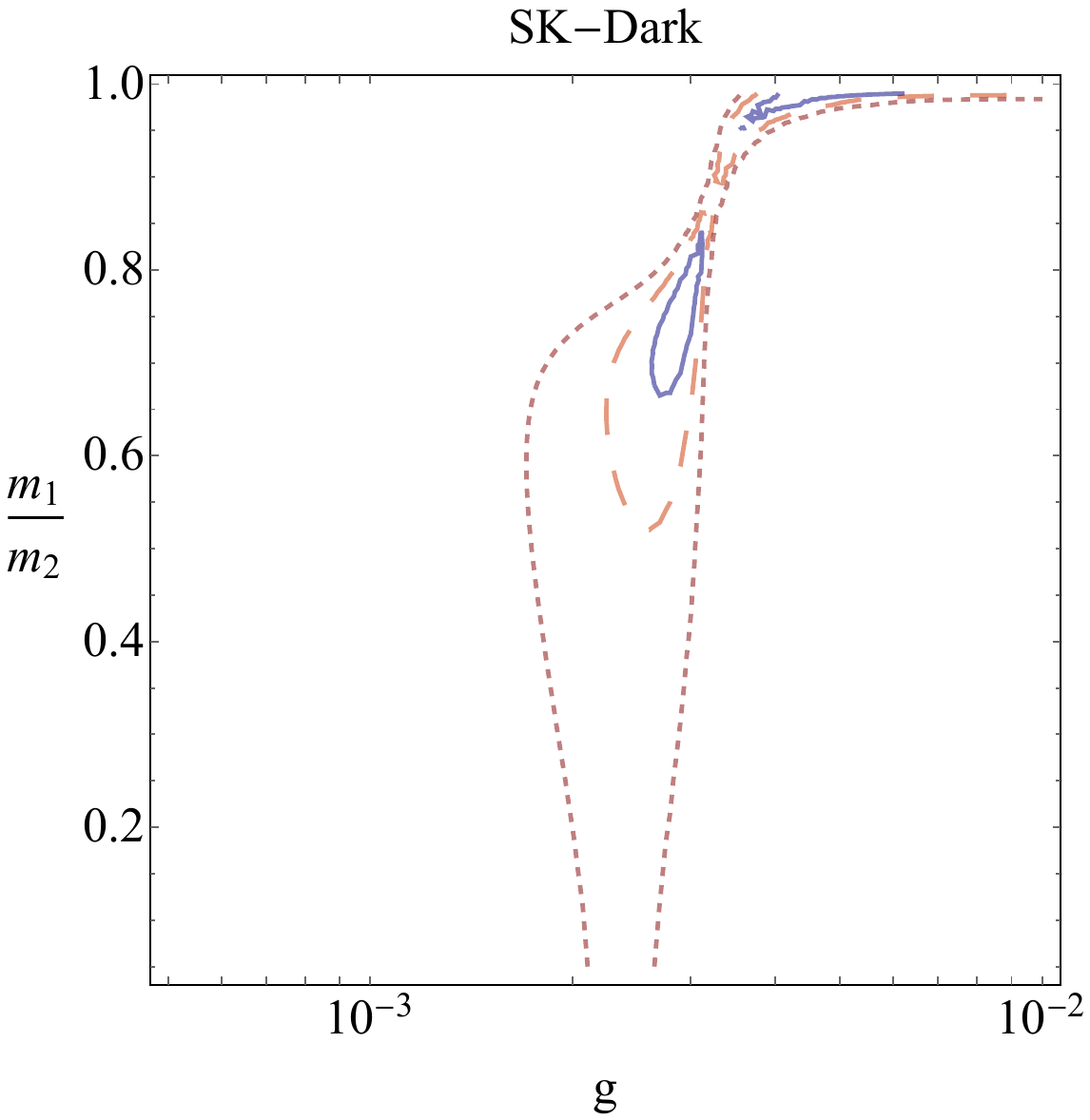}
    \caption{
    \label{fig:SK_Contourvisdec_md} Regions of the $g\times r$, $r=m_1/m_2$, parameter space allowed by Super-Kamiokande data assuming that external data constrain $\sin^2\theta_{12}=0.30\pm0.05$ (left) or $\sin^2\theta_{12}=0.70\pm0.05$ (right). See text for the details. The different contours correspond to one $\sigma$ or $\Delta \chi^2=2.30$ (solid), two $\sigma$ or $\Delta \chi^2=6.18$ (big dashed), and three $\sigma$ or $\Delta \chi^2=11.83$ (small dashed).}
\end{figure}

Fig.\,\ref{fig:SK_Contourvisdec_md} depicts the sensitivity of SK to $\nu_2$ decays in the $g\times r$ plane. We simulate SK data consistent with no decay, using the values of the mixing parameters listed in Eq.~(\ref{eq:osc_param}), and test for the impact of visible neutrino decays for different values of $g$ and $r$. We consider fifteen 0.5~MeV-wide bins of recoil-electron kinetic energy starting at 7~MeV and the equivalent of 504 days of data-taking, consistent with SK Phase-I. All oscillation parameters are kept fixed in this analysis, except for $\sin^2\theta_{12}$. As in the Borexino discussion, we impose external priors on $\sin^2\theta_{12}$. On the left-hand panel we impose $\sin^2\theta_{12}=0.3\pm0.05$ while the right-hand panel shows the result of a similar analysis where we instead choose a dark-side prior for the solar angle, $\sin^2\theta_{12}=0.7\pm0.05$. There is one very important distinction here: in the dark-side scenario (right-hand panel), $\Delta\chi^2$ is computed relative to the minimum value of $\chi^2$, $\chi^2_{\rm min}$, obtained in the light side. This is due to the fact that, in the no-decay scenario, the dark side is safely excluded by SK data and we wanted to ensure that the results here reflect the difference between the decay and no-decay hypotheses. The three contours correspond to one, two, and three $\sigma$ ($\Delta \chi^2=2.30$ (solid), $\Delta \chi^2=6.18$ (big dashed), and $\Delta \chi^2=11.83$ (small dashed), respectively).  

We first discuss the results presented in the left-hand panel of Fig.\,\ref{fig:SK_Contourvisdec_md}. As expected, there is no sensitivity to $g\lesssim 10^{-3}$; in this region of the parameter space, the lifetime is too long relative to the Earth--Sun distance. There is no allowed scenario in the region of parameter space where all $\nu_2$ decay into $\nu_1$ ($g\gg 0.001$). The reasoning is, rather qualitatively, as follows. When the decays are invisible (small $r$) one expects too few events in SK. Instead, when the decay is 100\% visible, the survival probability is too large (of order $\cos^2\theta_{12}\sim 0.7$). This implies there is a value of $r$ where the visible branching ratio is optimal. For such a value, however, the energy spectrum is distorted enough that a fit comparable to that of the stable $\nu_2$ case does not exist.

When $\sin^2\theta_{12}$ is constrained to the dark side, the situation is qualitatively different. In this case, when the lifetime is too long ($g\ll 0.001$), $\sin^2\theta_{12}\sim 0.7$ is safely ruled out since it leads to too many events associated to $^8$B neutrinos. For small $r$, large values of $g$ are also excluded. In this case, the $\nu_2$ decays invisibly and there are too few events  associated to $^8$B neutrinos. There is a range of values of $g$ that lead to a reasonable fit when $r$ is small. When $r$ is large and the decays are mostly visible, the situation is again qualitatively different. In the limit $r\to 1$ and large $g$, the $\nu_2$ population decays into left-handed $\nu_1$ with the same energy and, roughly, $P_{ee}\sim \cos^2\theta_{12}$, In the dark side, $\cos^2\theta_{12}\sim 0.3$ and we expect a fit that is just as good as the standard, no-decay fit when $\sin^2\theta_{12}=0.3$.   This corresponds to the region of parameter space allowed at one-sigma when $g$ is large and $r\to 1$.

\subsection{Sudbury Neutrino Observatory}
\label{sec:SNO}
The Sudbury Neutrino Observatory (SNO) was an underground neutrino detector in Canada, consisting of 1kT of heavy water (D$_2$O), designed to observe solar neutrinos. With heavy water, SNO could measure neutrinos through three channels:
\begin{enumerate}
    \item Charged current  (CC) interaction, $\nu_e + d \rightarrow p + p + e^-$, sensitive only to $\nu_e$,
   \item Neutral current(NC) interaction, $\nu_\alpha + d \rightarrow n + p + \nu_{\alpha}$, which is sensitive to neutrinos of all flavors, and 
   \item Elastic scattering on electrons (ES), $\nu_\alpha + e^- \rightarrow \nu_\alpha + e^-$, which is more sensitive to $\nu_e$ (by a factor of 5 or so) than to other flavors.
\end{enumerate}
The CC interaction provided a direct measurement of the $\nu_e$ flux from $^8$B neutrinos coming from the Sun, while the NC processes could measure the net $^8$B neutrino flux, irrespective of the flavor conversions. The impact of neutrino decay in SNO is similar to that in SK, with a few extra ingredients. The NC measurement is insensitive to the neutrino flavor but is sensitive to the total number of left-handed helicity neutrinos. Hence, invisible neutrino decays are constrained in a way that cannot be compensated by modifying the mixing parameters. On the other hand, visible neutrino decays allow one to accommodate the NC sample as long as the energy distribution of the daughter neutrinos is not very different from that of the parents. The CC measurement, instead, is only sensitive to electron-type neutrinos and hence it is impacted by neutrino decay in a way that is slightly different from the ES sample. Quantitatively, there is significantly more statistical power in the CC sample so we concentrate on it henceforth.

To simulate events in SNO, we consider 1 year of data taking, corresponding to roughly the first two phases of SNO, and consider the CC channel only.  We considered fourteen bins of equivalent-recoil-electron kinetic energy, 0.5~MeV wide \cite{SNO:2011hxd}. The number of targets is taken to be $N_{\rm tar}=3\times 10^{31}$. The resolution function is considered to be 
$\mathcal{S}(E_{\rm tr})=-0.462 + 0.5470\sqrt{E_{\rm tr} } + 0.00872 E_{\rm tr}$, following \cite{SNO:2011hxd} and the energy threshold of the detector is taken to be $E_{\rm thr}=1.446\,$MeV.
SNO reports the recoil kinetic energy $T_{\rm eff} =E_e - m_e$ and information is extracted from $T_{\rm eff}>6\,{\rm MeV}$. For lower energies, the NC sample dominates -- see Fig. 2 of \cite{SNO:2006odc}. 

\begin{figure}[!t]
    \centering
    \begin{tabular}{ccc}
        \includegraphics[width=0.32\textwidth]{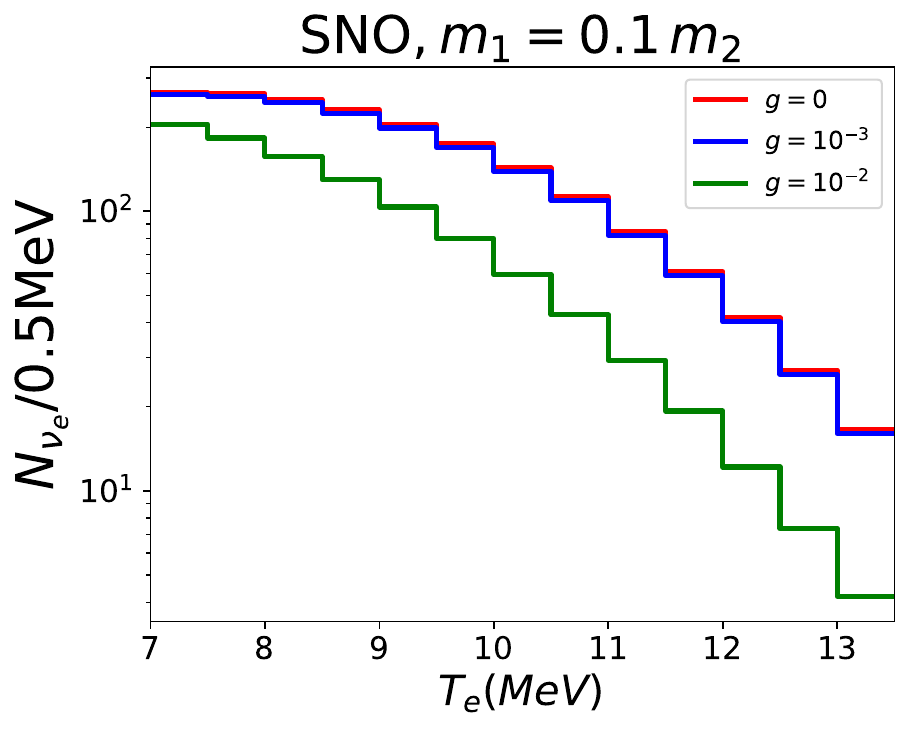}~
        \includegraphics[width=0.32\textwidth]{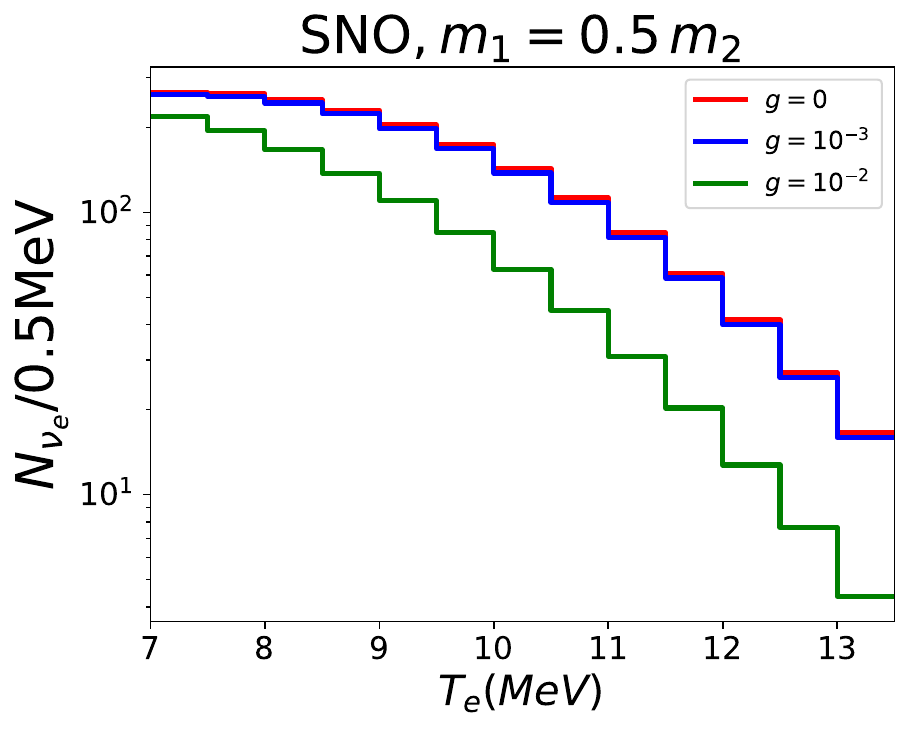}~
        \includegraphics[width=0.32\textwidth]{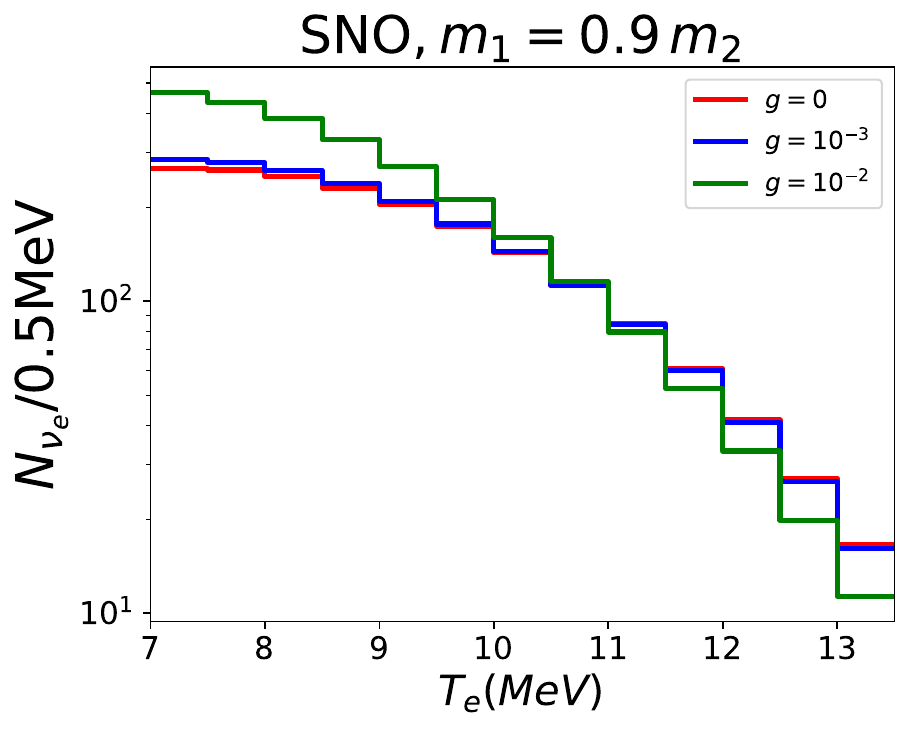}
    \end{tabular}
    \caption{
    \label{fig:SNO_visdec_md} Simulated SNO CC data for different values of the neutrino decay coupling $g$ and the ratio of the parent to daughter neutrino mass, $r=m_1/m_2$: $r=0.1$ (left), $r=0.5$ (center), and $r=0.9$ (right). $g=0$ corresponds to stable neutrinos, the standard oscillation scenario. All oscillation parameters are set to the values listed in Eq.~(\ref{eq:osc_param}).}
\end{figure}

\begin{figure}[!h]
 \centering
        \includegraphics[width=0.4\textwidth]{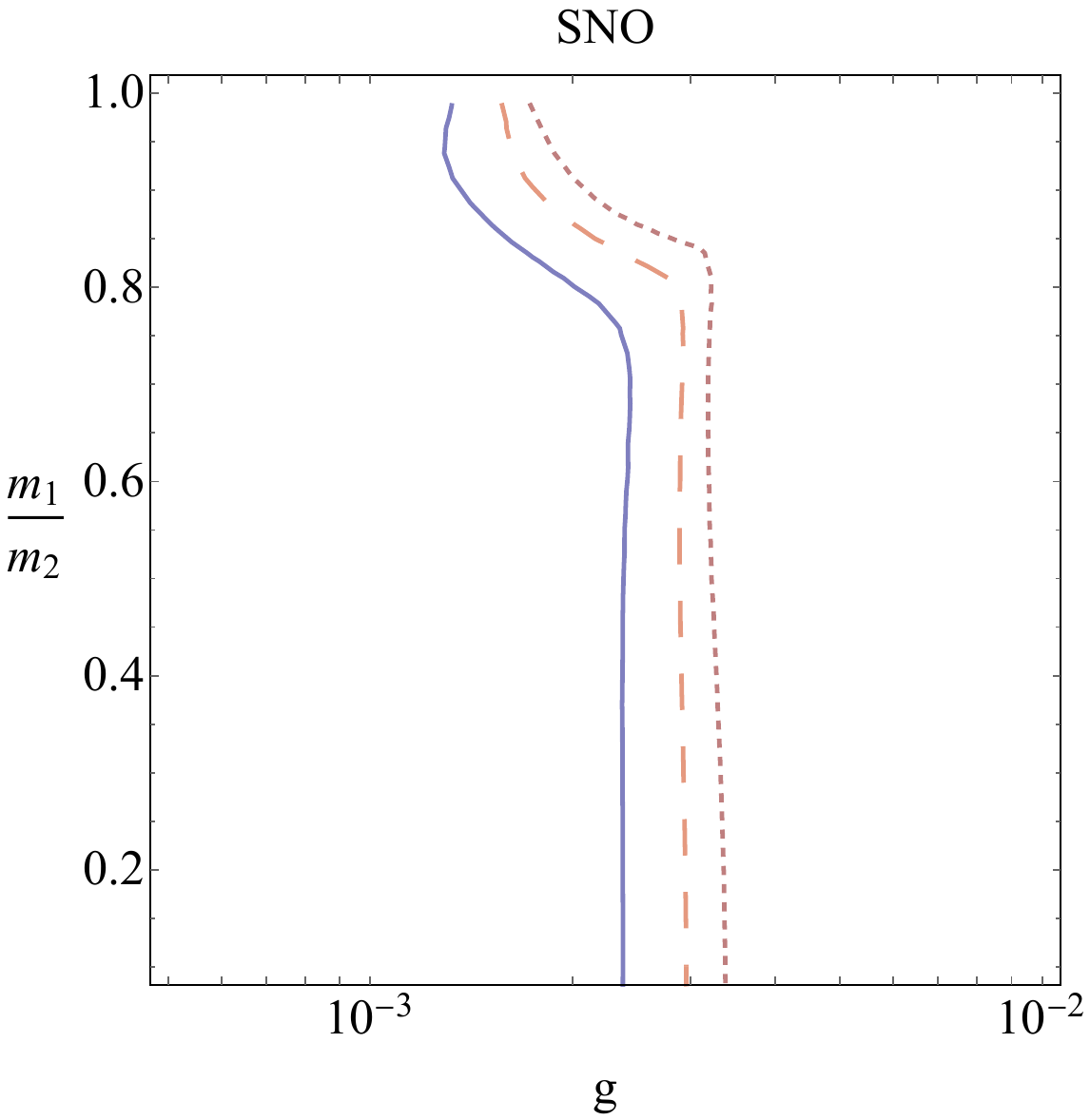}~~
         \includegraphics[width=0.4\textwidth]{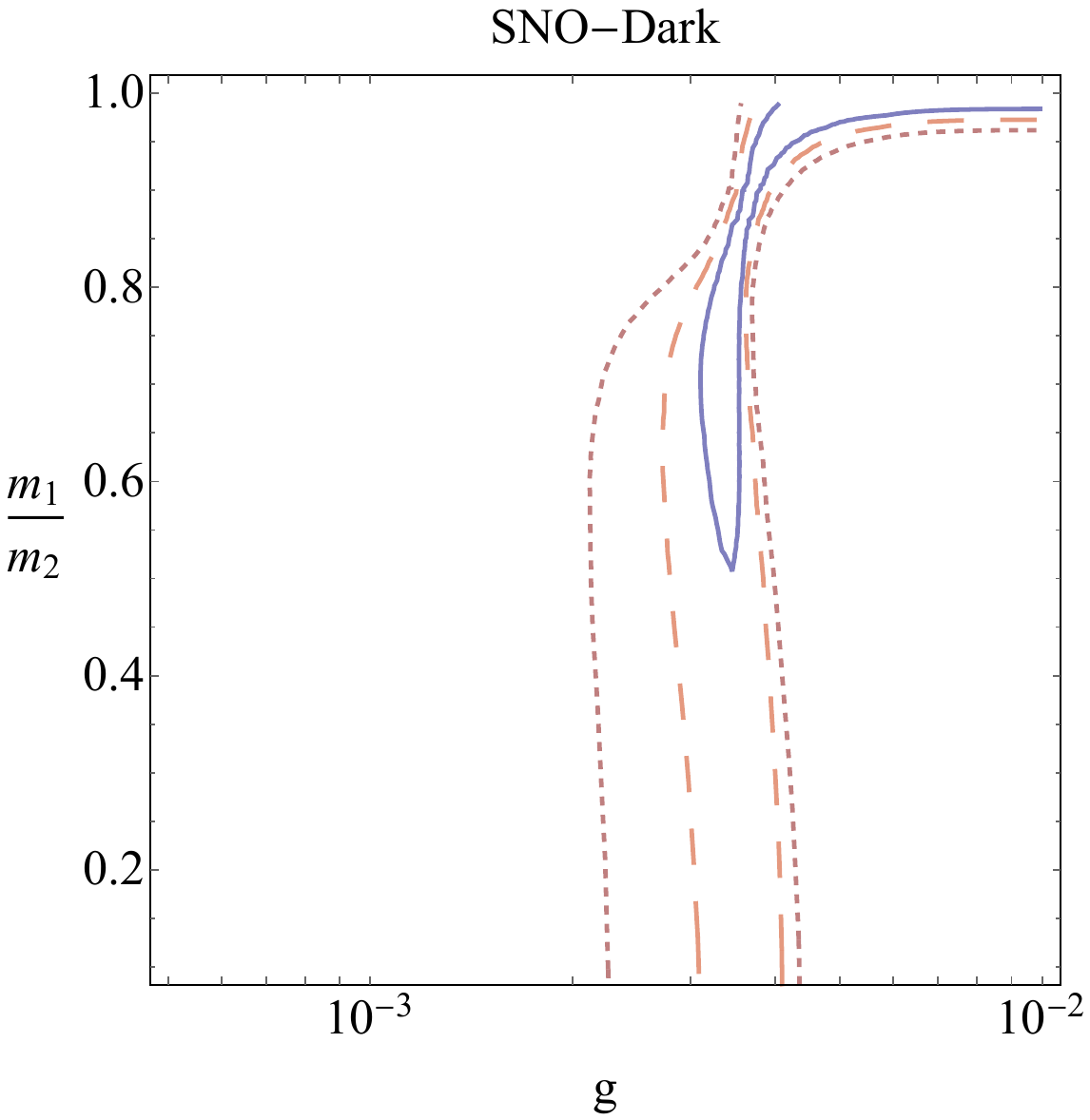}
    \caption{
    \label{fig:SNO_Contourvisdec_md} Regions of the $g\times r$, $r=m_1/m_2$, parameter space allowed by the SNO CC data assuming that external data constrain $\sin^2\theta_{12}=0.30\pm0.05$ (left) or $\sin^2\theta_{12}=0.70\pm0.05$ (right). See text for the details. The different contours correspond to one $\sigma$ or $\Delta \chi^2=2.30$ (solid), two $\sigma$ or $\Delta \chi^2=6.18$ (big dashed), and three $\sigma$ or $\Delta \chi^2=11.83$ (small dashed). }
\end{figure}

Fig.\,\ref{fig:SNO_visdec_md} shows the effect of visible decay on the CC event spectrum in SNO for different values of $g$ and $r$. We find features very similar to those at SK. Fig.\,\ref{fig:SNO_Contourvisdec_md} shows the sensitivity of the SNO CC sample to neutrino decays as a function of the coupling $g$ and the ratio of the daughter-parent masses $r\equiv m_1/m_2$. We simulate data consistent with no decay, and test for the impact of visible neutrino decays, as we vary $g$ and $r$. We find  results similar to those of SK, Fig.~\ref{fig:SK_Contourvisdec_md}. 

\subsection{Combined Results}
\label{sec:combined}
\begin{figure}[htb]
\centering
\includegraphics[width=0.4\textwidth]{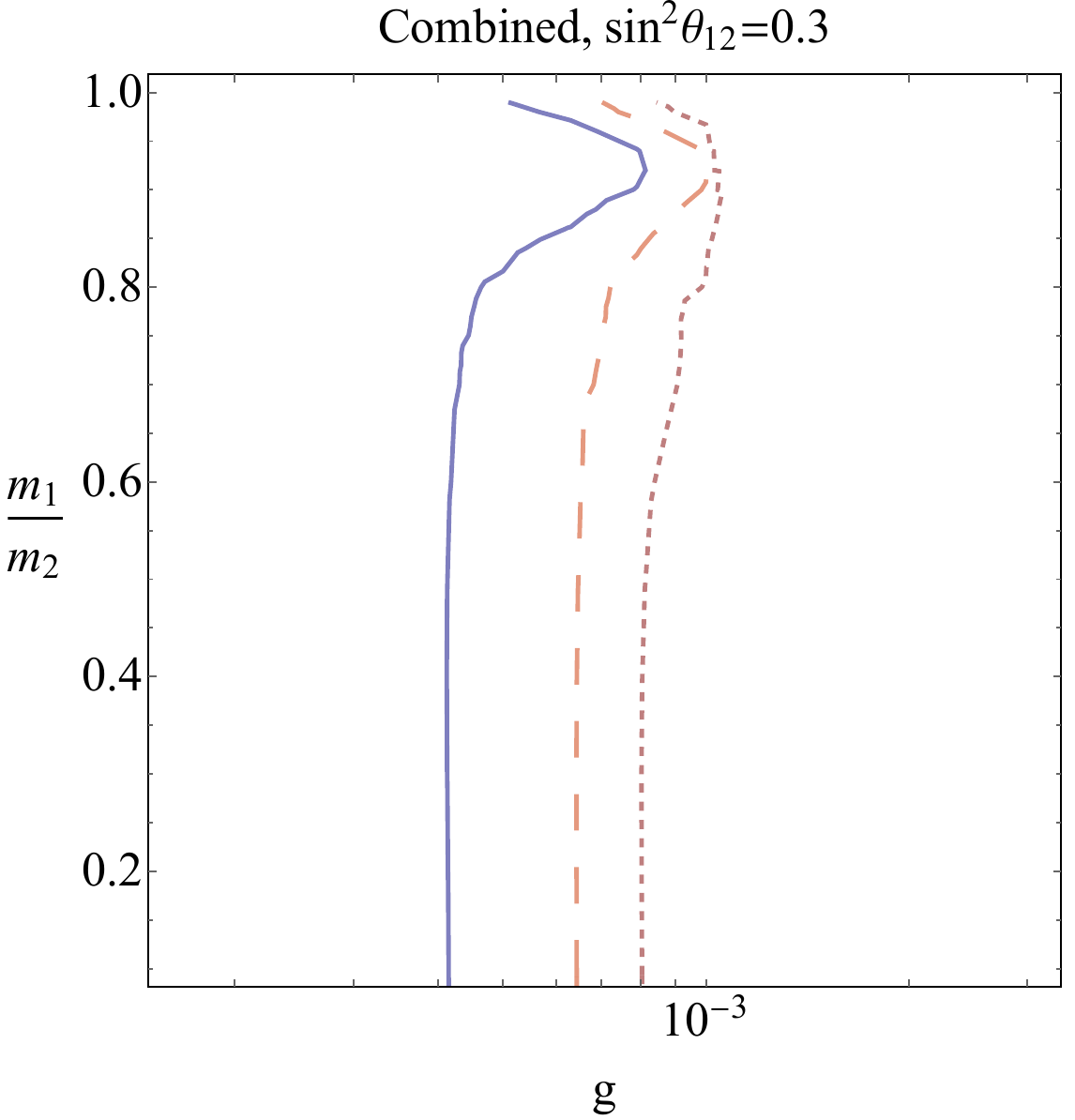}
\caption{\label{fig:Comb} Regions of the $g\times r$, $r=m_1/m_2$, parameter space allowed by the combination of Borexino, SK, and SNO data assuming that external data constrain $\sin^2\theta_{12}=0.30\pm0.05$. See text for the details. The different contours correspond to one $\sigma$ or $\Delta \chi^2=2.30$ (solid), two $\sigma$ or $\Delta \chi^2=6.18$ (big dashed), and three $\sigma$ or $\Delta \chi^2=11.83$ (small dashed).}
\end{figure}

In this section we describe our combined analysis for the three experiments. The light-side result is presented in Fig.\,\ref{fig:Comb} and shows the combined $\Delta\chi^2$ contour plot for SK, SNO, and Borexino for decay relative to no decay with a prior of $\sin^2\theta_{12}=0.30\pm 0.05$.  Contours are for $\Delta \chi^2=2.30$ (solid), $\Delta \chi^2=6.18$ (big dashed), and $\Delta \chi^2=11.83$ (small dashed). We repeated the same exercise with a dark-side prior, $\sin^2\theta_{12}=0.70\pm 0.05$, where, similar to Figs.~\ref{fig:SK_Contourvisdec_md} and \ref{fig:SNO_Contourvisdec_md}, $\Delta\chi^2$ is computed relative to the minimum value of $\chi^2$, $\chi^2_{\rm min}$, obtained in the light side. In this case, we find a very small allowed region -- too small to capture in a useful figure -- characterized by $g\sim 0.005$ and $r\sim 0.98$. There are, however, points in the parameter space allowed at close to the one-sigma level. 

Independent from the assumption on $\sin^2\theta_{12}$ -- light side versus dark side -- the complementarity between the low energy solar neutrino data (Borexino) and the high energy solar neutrino data (SK and SNO) is clear. The reason is the mass-eigenstate composition of the $^7$Be solar neutrino flux is quite different from that of the $^8$B: the $^7$Be flux is almost 70\% $\nu_1$ while the $^8$B flux is more than 90\% $\nu_2$. In the light side, Borexino data allow for short lifetimes as long as the ratio of $m_1$ and $m_2$ is around 0.9 but this possibility is safely excluded by SK and SNO data. In the dark side, instead, SK and SNO data allow for short lifetimes as long as the ratio of $m_1$ and $m_2$ is close to 1. This is disfavored by Borexino data. In the end, the combined data sets allow $\sin^2\theta_{12}$ around 0.7 as long as $m_1$ and $m_2$ are close in mass and $\nu_2$ decays to $\nu_1$ between the Earth and the Sun. This hypothesis is only slightly disfavored and is certainly not excluded by our rather simplified analyses.

\section{Concluding Remarks}
\label{sec:conclusions}

We explored how active neutrino decays impact the interpretation of solar neutrino data and how well solar neutrino data can test the hypothesis that the active neutrinos are unstable. We were especially interested in understanding the nontrivial impact of nonzero daughter neutrino masses in the different analyses. For the scenario of interest, we found that the value of the daughter neutrino mass can significantly modify the impact of active solar neutrino decay, sometimes qualitatively. We also explored the possibility that the solar angle resides in the so-called dark side of the parameter space, a hypothesis that, for stable neutrinos, is only excluded by solar neutrino data. It is important to revisit this possibility whenever one modifies the physics of solar neutrinos. It is well known, for example, that the addition of large non-standard neutrino--matter interactions allow a fit to the solar neutrino data in the dark side \cite{Miranda:2004nb}. We found that measurements of $^7$Be and $^8$B neutrinos are quite complementary. ``Blind spots'' in one type of experiment are often covered by the other type. Nonetheless, we find that the hypothesis that the solar angle resides on the dark side is allowed as long as $\nu_1$ and $\nu_2$ are close in mass and $\nu_2$ decays relatively quickly relative to the Earth--Sun distance. A more sophisticated analysis, outside the aspirations of this manuscript, is necessary in order to reveal whether a dark-side solution with unstable $\nu_2$ is indeed allowed by the combined SK, SNO, and Borexino data. We highlight the importance of Borexino when it comes to addressing this particular issue.  

We restricted our discussions to the hypothesis that neutrinos are Dirac fermions and that the decay is governed by Eq.~(\ref{eq:Lagrangian}), for a few reasons. We chose a chiral interaction such that, when the daughter neutrino is massless, it has right-handed helicity and is hence ``sterile.'' The situation is very different when the mass of the daughter neutrino approaches that of the parent. This way, we can change the nature of the decay by dialing up or down the daughter neutrino mass. We also chose a very simple model -- a two body decay -- in order to render our results and discussions as transparent as possible. Finally, with Dirac neutrinos, we did not have to worry about the possibility of ``neutrinos'' converting into ``antineutrinos.'' 

There is a price to be paid by our choice of decay Lagrangian.  Eq.~(\ref{eq:Lagrangian}) is not $SU(2)_L\times U(1)_Y$ invariant, for example. A more complete version of this model would include new interactions involving charged leptons, gauge bosons, or other hypothetical new particles. Furthermore, the same physics that mediates neutrino decay will also mediate other phenomena that will provide more nontrivial constraints on the new-physics coupling $g$. These include relatively long-range neutrino--neutrino interactions -- see \cite{Berryman:2022hds} for a recent thorough overview -- the presence of new light degrees of freedom in the early universe -- see, for example, \cite{Venzor:2020ova,Taule:2022jrz, Das:2023npl, Sandner:2023ptm} for recent analyses -- and low-energy laboratory processes -- see, for example, \cite{Pasquini:2015fjv,Farzan:2018gtr,Bickendorf:2022buy}. We did not take any of these constraints into account here. 

On the other hand, the results discussed here can be generalized to other interesting decay scenarios. As introduced and discussed in \cite{deGouvea:2022cmo}, for Dirac neutrinos, neutrino decay can be mediated by a four-fermion interaction that involves only right-chiral neutrino fields (see Eq.~(II.5) in \cite{deGouvea:2022cmo}). New interactions that involve only gauge-singlet fermions are virtually unconstrained by experiments and observations, and are probably best constrained by searches for active neutrino decays. When the neutrino mass ordering is inverted, the decay $\nu_2\to\nu_1\nu_3\bar{\nu}_3$, in the limit where $m_2$ and $m_1$ are quasi-degenerate and the mass $m_3$ of the lightest neutrino is very small relative to $m_1$,\footnote{As discussed earlier, $m_1/m_2$ very close to one is guaranteed in the case of the inverted mass ordering. Furthermore, in the limit $m_3\ll m_2$, the $\nu_3$ and $\bar{\nu}_3$ are, for all practical purposes, both ``sterile.''} is kinematically quite similar to the decay $\nu_2\to\nu_1\varphi$. The results obtained here can be adapted to this other decay scenario without too much difficulty.

\section*{Acknowledgements}
We thank Pedro de Holanda, Orlando Peres, Renan Picoreti, and Dipyaman Pramanik for helpful discussions and collaboration in the initial stages of the project. This work was supported in part by the US Department of Energy (DOE) grant \#de-sc0010143 and in part by the NSF grant PHY-1630782. AdG also acknowledges the warm hospitality of the Department of Theoretical Physics and Cosmology at the University of Granada, where some of this work was carried out. 


\bibliographystyle{kpmod}
\bibliography{DecayNu}

\end{document}